\documentclass[aps,showpacs,floatfix]{revtex4}

\usepackage{amssymb}
\usepackage{amsmath}
\usepackage{epsfig}
\usepackage{subfigure}
\usepackage{multirow}

\begin{document}

\title{Decaying Gravity}
\author{T. Clifton}
\email{T.Clifton@damtp.cam.ac.uk}
\affiliation{DAMTP, Centre for Mathematical Sciences, University of Cambridge,
Wilberforce Road, Cambridge, CB3 0WA, UK}
\author{John D. Barrow}
\email{J.D.Barrow@damtp.cam.ac.uk}
\affiliation{DAMTP, Centre for Mathematical Sciences, University of Cambridge,
Wilberforce Road, Cambridge, CB3 0WA, UK}
\date{\today }

\begin{abstract}
We consider the possibility of energy being exchanged between the scalar and
matter fields in scalar-tensor theories of gravity. Such an exchange
provides a new mechanism which can drive variations in the gravitational
`constant' $G$. We find exact solutions for the evolution of spatially flat
Friedman-Robertson-Walker cosmologies in this scenario and discuss their
behaviour at both early and late times.  We also consider the physical
consequences and observational constraints on these models.
\end{abstract}

\maketitle

\section{Introduction}

Scalar-tensor theories of gravity provide a convenient framework within
which to model space-time variations of the Newtonian gravitational
constant, $G$. They feature a scalar field, $\phi $, which is non-minimally
coupled to the space-time curvature in the gravitational action. It is this
scalar, or more usually its reciprocal, that drives variations in $G$.
Non-minimally coupled scalar fields arise in a variety of different
theories, including Kaluza-Klein theory \cite{KK}, string theories \cite%
{string} and brane-worlds \cite{branes}. The same mechanism that creates a
scalar field non-minimally coupled to the curvature in these theories can
also lead to a coupling between the scalar and matter fields. This coupling
manifests itself through the matter Lagrangian becoming a function of $\phi $%
. The possibility of such a coupling is usually neglected in the literature,
where the matter Lagrangian is \textit{a priori} assumed to be independent
of $\phi $. It is the possibility of a coupling between the scalar and
matter fields in scalar-tensor theories that will be the subject of this
work.

The introduction of a coupling between $\phi $ and matter greatly enlarges
the phenomenology of the theory. Potentially, this allows greater
variability of $G$ in the early universe whilst still satisfying the solar
system bounds on time-varying $G$ \cite{Bert}. We will consider
spatially-flat Friedmann-Robertson-Walker (FRW) cosmologies and investigate
the extent to which $G$ can vary when energy is exchanged between the $\phi $
field and ordinary matter. As well as giving a window into the
four-dimensional cosmologies associated with higher-dimensional theories, we
hope that this direction of study might also be useful in understanding why
the present value of $G$ is so small compared to the proton mass scale ($%
Gm_{pr}^{2}\sim 10^{-39}$). The direct exchange of energy between $\phi $
and the matter fields offers a non-adiabatic mechanism for $G$ to `decay'
towards its present value from a potentially different initial value. There
have been a variety of studies which investigate the drain of energy from
ordered motion by entropy generation, due to bulk viscosity \cite{bulk} or
direct decay \cite{Bar04, fr} or energy exchange \cite{tolman}, but few
studies of the drain of energy by non-adiabatic processes from a scalar
field that defines the strength of gravity \cite{p1,p2,p3}. This creates a range of new
behaviours in scalar-tensor cosmologies.

In considering a coupling between $\phi $ and matter we are forced to
reconsider the equivalence principle. The energy-momentum tensor of
perfect-fluid matter fields will no longer be covariantly conserved and the
trajectories of test-particles will no longer follow exact geodesics of the
metric. These violations of the experimentally well verified weak
equivalence principle exclude most possible couplings between $\phi $ and
matter \cite{will}. Such violations are not necessarily fatal though. We
show that whilst energy-momentum is not separately conserved by the matter
fields there is still an exact concept of energy-momentum conservation when
the energy density of the scalar field is included. Furthermore, the
non-geodesic motion of test particles is only problematic if the coupling
increases above experimentally acceptable levels as the Universe ages. The
theory we consider is still a geometric one and it remains true that at any
point on the space-time manifold it is possible to choose normal coordinates
so that it looks locally flat, ensuring that it is always possible to
transform to a freely-falling frame in which the effects of gravity are
negligible (up to tidal forces).

\section{Field Equations}

The simplest scalar-tensor theory is the Brans-Dicke theory \cite{bd},
defined by the Lagrangian density 
\begin{equation}
\mathcal{L}=\phi R-\frac{\omega }{\phi }\partial ^{a}\phi \partial _{a}\phi
+16\pi \mathcal{L}_{m}[g_{ab};\Psi ]  \label{BD}
\end{equation}%
where $\omega $ is the Brans-Dicke coupling constant, $R$ is the scalar
curvature of space-time, and $\mathcal{L}_{m}$ is the Lagrangian density of
the matter fields, denoted by $\Psi $. As $\omega \rightarrow \infty $ this
theory reduces to general relativity and $G\sim \phi ^{-1}$ becomes
constant (for exceptions see \cite{Rom, Ban}). It is an important feature of these theories that $\mathcal{L}_{m}$
is independent of $\phi $. This ensures that the matter fields do not
interact with the scalar field directly and therefore that the
energy-momentum tensor, $T_{{}}^{ab}$, derived from $\mathcal{L}_{m}$ is
conserved (${T_{{}}^{ab}}_{;b}=0$).

This conservation of $T_{{}}^{ab}$, whilst appealing, is not absolutely
necessary in deriving a theory in which $G$ can vary. There are numerous
examples where one might expect ${T_{{}}^{ab}}_{;b}\neq 0$. For example,
when considering two fluids the energy-momentum tensor of each fluid is not
separately conserved unless the fluids are completely non-interacting. It is
only required that the energy-momentum being lost by one of the fluids is
equal to the energy-momentum being gained by the other.

In what follows we will consider the scalar field and the matter fields as
two fluids (or more than two fluids if there is more than one matter fluid
present) and introduce a transfer of energy and momentum between them. Such
an interaction can be introduced by allowing $\mathcal{L}_{m}$ to be a
function of $\phi $ and will change the nature of the resulting FRW
cosmologies.

The field equations are derived from (\ref{BD}) by extremizing the
corresponding action with respect to the metric. Defining $T_{{}}^{ab}$ for
the matter in the usual way, the field equations take their standard
Brans-Dicke form \cite{bd} independent of the presence of interactions
between $\phi $ and matter, and the Einstein tensor is given by 
\begin{equation}
G^{ab}=\frac{\omega }{\phi ^{2}}({\phi _{;}}^{a}{\phi _{;}}^{b}-\frac{1}{2}%
g^{ab}\phi _{;c}{\phi _{;}}^{c})+\frac{1}{\phi }({\phi _{;}}%
^{ab}-g^{ab}\square \phi )+\frac{8\pi }{\phi }T_{m}^{ab}.  \label{field}
\end{equation}%
The scalar-field propagation equation and matter energy-momentum
conservation equations are 
\begin{align}
\square \phi & =\frac{8\pi T}{(2\omega +3)}-\frac{16\pi \phi }{(2\omega +3)}%
\frac{\sigma ^{a}}{{\phi _{;}}^{a}}  \label{field2} \\
{T^{ab}}_{;b}& =\sigma ^{a}  \label{field3}
\end{align}%
where $T$ is the trace of the energy-momentum tensor and $\sigma ^{a}$ is an
arbitrary vector function of the space-time coordinates $x^{b}$ that
determines the rate of transfer of energy and momentum between the scalar
field $\phi $ and the ordinary matter fields. The precise form of $\sigma
^{a}$ depends on the detailed form of the interaction between the scalar and
matter fields in $\mathcal{L}_{m}$. For example, a conformal transformation
of the form $g_{ab}\rightarrow A^{2}(\phi )g_{ab}$ from a frame in which ${%
T_{\ \ ;b}^{ab}}=0$ gives 
\begin{equation*}
\sigma ^{a}=\frac{T}{A}\frac{dA}{d\phi }{\phi _{;}}^{a}.
\end{equation*}%
This particular choice of energy transfer can be interpreted as a space-time
variation of the rest masses of matter described by $\mathcal{L}_{m}$. For
the moment, we consider the case of more general interactions by leaving $%
\sigma ^{a}$ as an arbitrary function. Later, we will consider specific
forms of $\sigma ^{a}$ that allow direct integration of the field equations.

We specialise the metric to the spatially flat, isotropic and homogeneous
FRW line-element with expansion scale factor $a(t)$: 
\begin{equation}
ds^{2}=-dt^{2}+a^{2}(t)(dr^{2}+r^{2}(d\theta ^{2}+\sin ^{2}\theta d\phi
^{2})).  \label{FRW}
\end{equation}%
Substituting this metric into the field equations (\ref{field}), (\ref%
{field2}) and (\ref{field3}) gives the generalised Friedmann equations: 
\begin{align}
\left( \frac{\dot{V}}{V}\right) ^{2}& =-3\frac{\dot{V}}{V}\frac{\dot{\phi}}{%
\phi }+\frac{3\omega }{2}\left( \frac{\dot{\phi}}{\phi }\right)
^{2}+3(3+2\omega )\alpha \frac{\rho }{\phi }  \label{1} \\
\frac{(\dot{V}\phi )^{\cdot }}{\rho V}& =3\alpha ((2-\gamma )\omega +1)+3%
\frac{\alpha }{\rho }\frac{\phi }{\dot{\phi}}\sigma ^{0}  \label{2} \\
\frac{(V\dot{\phi})^{\cdot }}{\rho V}& =\alpha (4-3\gamma )-2\frac{\alpha }{%
\rho }\frac{\phi }{\dot{\phi}}\sigma ^{0}  \label{3} \\
\dot{\rho}+\gamma \frac{\dot{V}}{V}\rho & =\sigma ^{0}  \label{4}
\end{align}%
where we have defined a comoving volume $V=a^{3}$ and a constant $\alpha =%
\frac{8\pi }{(3+2\omega )}$; the energy-momentum tensor is assumed to be a
perfect barotropic fluid with density $\rho $ and pressure $p$ which are
linked by a linear equation of state $p=(\gamma -1)\rho $, and over-dots
denote differentiation with respect to the comoving proper time, $t$. It is
this set of differential equations that we need to solve in order to
determine the evolution of $a(t)$ and $G\propto \phi (t)^{-1}$ in
cosmological models of this type.

\section{Transfer of Energy and Entropy}

The conservation of energy and momentum as well as the second law of
thermodynamics are of basic importance to physics.  In considering an
interaction between a gravitational scalar field $\phi$ and
the matter fields $\Psi$ it is, therefore, necessary to investigate the extent to which
we can consider energy and momentum to be conserved and the second law
to be obeyed.

When we consider the thermodynamics of an exchange of energy between the
scalar field and matter it is useful to define an effective energy density, $%
\rho _{\phi }$, for the scalar field $\phi $. Defining 
\begin{equation}
\rho _{\phi }\equiv \frac{\dot{\phi}^{2}}{16\pi \omega \phi },  \label{r1}
\end{equation}%
the scalar-field propagation equation (\ref{2}) can then be rewritten as 
\begin{equation}
\dot{\rho}_{\phi }+2\frac{\dot{V}}{V}\rho _{\phi }=-\frac{R}{16\pi }\dot{\phi%
}-\sigma ^{0}.  \label{r2}
\end{equation}%
Comparison of this equation with (\ref{4}) shows that $\phi $ acts as a
fluid with equation of state $\gamma =2$ ($p_{\phi }=\rho _{\phi }$). The
two terms on the right hand side of this equation act as sources for the
energy density $\rho _{\phi }$. The first is the standard Brans-Dicke source
term for the scalar field and the second, $\sigma ^{0}(t)$, is new and
describes the energy exchange between $\phi $ and the matter fields. It can
be seen that the second term is exactly the opposite of the source term in
equation (\ref{1}), and it is in this sense that the total energy is
conserved in this theory.

It is also useful to consider the entropy. Contracting the divergence of the
energy-momentum tensor with the comoving four-velocity $U^{a}$ we obtain 
\begin{align*}
U^{a}\sigma _{a}& =U^{a}{{T_{a}}^{b}}_{;b} \\
& =U^{a}p_{;a}+U^{a}((\rho +p)U_{a}U^{b})_{;b} \\
& =U^{a}p_{;a}-((\rho +p)U^{b})_{:b},
\end{align*}%
where, in the last line, we have used the normalisation $U^{a}U_{a}=-1$ and $%
(U^{a}U_{a})_{;b}={U^{a}}_{;b}U_{a}+U^{a}{U_{a}}_{;b}=0$. Defining the
particle current\ by $N^{a}\equiv nU^{a}$, where $n$ is the number density
in a comoving Lorentz frame, this expression can be rewritten as 
\begin{align*}
U^{a}\sigma _{a}& =U^{a}\left[ p_{;a}-n\left( \frac{(\rho +p)}{n}\right)
{}_{;a}\right] -\frac{(\rho +p)}{n}{N^{a}}_{;a} \\
& =-nU^{a}\left[ p\left( \frac{1}{n}\right) {}_{;a}+\left( \frac{\rho }{n}%
\right) {}_{;a}\right] ,
\end{align*}%
where we have used the conservation of particle number, ${N^{a}}_{;a}=0$.
Recalling the first law of thermodynamics, 
\begin{equation*}
\Theta dS=pdV+dE=pd\left( \frac{1}{n}\right) +d\left( \frac{\rho }{n}\right)
,
\end{equation*}%
where $\Theta $ is the temperature and $S$ is the entropy, we now get 
\begin{equation*}
U^{a}\sigma _{a}=-n\Theta U^{a}S_{;a}
\end{equation*}%
or, making use of our assumption of spatial homogeneity, 
\begin{equation*}
\dot{S}=\frac{\sigma ^{0}}{n\Theta }.
\end{equation*}%
This tells us that as energy is transferred from $\phi $ to the matter
fields the entropy of the matter fields increases, as expected. Conversely,
the matter fields can decrease their entropy by transferring energy into $%
\phi $.

Unfortunately, there is currently no known way of defining
the entropy of a non-static gravitational field so it is not possible to perform
an explicit calculation of the entropy changes in $\phi$ and $g_{a b}$.  We can
only assume that if the Universe can be treated as a closed system, and
the exchange of energy is an equilibrium process, then the entropy
that is lost or gained by the matter through this exchange will be
gained or lost by the gravitational fields.  This direct interaction
of the matter with $\phi$ then allows an additional mechanism for
increasing or decreasing the entropy of the matter content of the Universe.

\section{General Solutions}

It is convenient to define a new time coordinate $\tau $ by

\begin{equation}
d\tau \equiv \rho Vdt  \label{tau}
\end{equation}
and to re-parametrise the arbitrary function $\sigma ^{0}$ by 
\begin{equation}
\sigma ^{0}=\rho ^{2}V\frac{\phi ^{\prime }}{\phi }\lambda ^{\prime }
\label{sig}
\end{equation}%
where a prime denotes differentiation with respect to $\tau $ and $\lambda
(\tau )$ is a new arbitrary function.  This re-parameterisation of the
interaction is chosen to
enable a direct integration of the field equations and does not imply
any loss of generality, as $\lambda$ is an arbitrary function.  The field
equations (\ref{2}) and (\ref{3}) can now be integrated to 
\begin{align}
\rho \phi VV^{\prime }& =3\alpha ((2-\gamma )\omega +1)(\tau -\tau
_{1})+3\alpha \lambda  \label{V'} \\
\rho V^{2}\phi ^{\prime }& =\alpha (4-3\gamma )\tau +\alpha \tau
_{2}-2\alpha \lambda  \label{phi'}
\end{align}%
where $\tau _{1}$ and $\tau _{2}$ are constants of integration. We have a
freedom in where we define the origin of $\tau $ and can, therefore, absorb
the constant $\tau _{1}$ into $\tau $ and the definition of $\tau _{2}$ by
the transformations $\tau \rightarrow \tau +\tau _{1}$ and $\tau
_{2}\rightarrow \tau _{2}-(4-3\gamma )\tau _{1}$. It can now be seen from (%
\ref{phi'}) that $\phi ^{\prime }$ is sourced by three terms. The first
corresponds to the source term in (\ref{field3}) and can be seen to
disappear for $\gamma =4/3$, as expected for black-body radiation. The
second term is constant and is the contribution of the free scalar to the
evolution of $\phi $; it is this term which distinguishes the general
spatially-flat Brans-Dicke FRW solutions \cite{Gur73} from the power-law
late-time attractor solutions \cite{Nar}. The third term is new and gives
the effect of the energy transfer on the evolution of $\phi $. This term is
dependent on the arbitrary function $\lambda $, which specifies the
interaction between $\phi $ and the matter fields.

The problem is now reduced to solving the coupled set of first-order
ordinary differential equations (\ref{V'}) and (\ref{phi'}) with the
constraint equation (\ref{4}). The remaining equation (\ref{1}) is rewritten
in terms of $\tau $ and $\lambda $ as 
\begin{equation}
\frac{\rho ^{\prime }}{\rho }+\gamma \frac{V^{\prime }}{V}=\lambda ^{\prime }%
\frac{\phi ^{\prime }}{\phi },  \label{rho}
\end{equation}%
and can be solved for $\rho$ once $V$ and $\phi $ have been found for some $%
\lambda $.

We can decouple the set of equations (\ref{V'}) and (\ref{phi'}) by
differentiating (\ref{phi'}) and substituting for (\ref{V'}) to get the
second-order ordinary differential equation 
\begin{align*}
& \left( (4-3\gamma )\tau +\tau _{2}-2\lambda \right) \frac{\phi ^{\prime
\prime }}{\phi }-\left( (4-3\gamma )-2\lambda ^{\prime }\right) \frac{\phi
^{\prime }}{\phi } \\
=& -\left[ ((4-3\gamma )\tau +\tau _{2}-2\lambda )\lambda ^{\prime
}+3(2-\gamma )(((2-\gamma )\omega +1)\tau +\lambda )\right] \left( \frac{%
\phi ^{\prime }}{\phi }\right) ^{2}
\end{align*}%
which can be integrated to 
\begin{equation}
\frac{\phi ^{\prime }}{\phi }=\frac{(4-3\gamma )\tau +\tau _{2}-2\lambda }{%
(A\tau ^{2}+B\tau +C)},  \label{phi'2}
\end{equation}%
where 
\begin{align*}
A& =3\gamma ^{2}\omega /2-3\gamma (1+2\omega )+(5+6\omega ) \\
B& =\tau _{2}+(4-3\gamma )\lambda \\
C& =-\lambda ^{2}+\lambda \tau _{2}+D
\end{align*}%
and $D$ is a constant of integration. The three source terms for $\phi
^{\prime }$ appear in the numerator on the right-hand side of equation (\ref%
{phi'2}). The equations (\ref{V'}) and (\ref{phi'}) can now be combined to
give $a^{\prime }/a$ in terms of $\phi ^{\prime }/\phi $ as 
\begin{equation}
\frac{a^{\prime }}{a}=\frac{((2-\gamma )\omega +1)\tau +\lambda }{(A\tau
^{2}+B\tau +C)}  \label{V'2}
\end{equation}%
where $A$, $B$, $C$ and $D$ are defined as before. The constant $D$ can be
set using the constraint equation (\ref{4}). Using (\ref{V'}) and (\ref{V'2}%
) we obtain the expression 
\begin{equation*}
\rho V^{2}\phi =\alpha (A\tau ^{2}+B\tau +C).
\end{equation*}%
This can then be substituted into (\ref{4}) which, in terms of $\tau $ and $%
a $, gives the generalised Friedmann equation: 
\begin{equation*}
3\left( \frac{a^{\prime }}{a}\right) ^{2}+3\frac{a^{\prime }}{a}\frac{\phi
^{\prime }}{\phi }-\frac{\omega }{2}\left( \frac{\phi ^{\prime }}{\phi }%
\right) ^{2}=\frac{(3+2\omega )}{(A\tau ^{2}+B\tau +C)}.
\end{equation*}%
Substituting (\ref{phi'2}) and (\ref{V'2}) into this we find that 
\begin{equation*}
D=-\frac{\tau _{2}^{2}\omega }{2(3+2\omega )}.
\end{equation*}

\section{Particular Solutions}

If $\lambda $ is specified in terms of $\tau $ we now have a set of two
decoupled first-order ordinary differential equations for the two variables $%
a$ and $\phi $. It is the solution of these equations, for specific choices
of $\lambda (\tau )$, that we give in this section.

\subsection{$\protect\lambda (\protect\tau)= c_1+c_2 \protect\tau$}

A simple form for $\lambda $ that allows direct integration of equations (%
\ref{phi'2}) and (\ref{V'2}) is the linear function $\lambda
=c_{1}+c_{2}\tau $. From equations (\ref{V'}) and (\ref{phi'}) it can be
seen that the constant $c_{1}$ can be absorbed into $\tau _{1}$ and $\tau
_{2}$ by simple redefinitions. The equations (\ref{phi'2}) and (\ref{V'2})
then become 
\begin{align*}
\frac{\phi ^{\prime }}{\phi }& =\frac{(4-3\gamma -2c_{2})\tau +\tau _{2}}{(%
\hat{A}\tau ^{2}+\hat{B}\tau +\hat{C})} \\
\frac{a^{\prime }}{a}& =\frac{((2-\gamma )\omega +1+c_{2})\tau }{(\hat{A}%
\tau ^{2}+\hat{B}\tau +\hat{C})}
\end{align*}%
where $\hat{A}=A-c_{2}^{2}$, $\hat{B}=B+c_{2}\tau _{2}$ and $\hat{C}=D$. The
solutions of these equations depend upon the sign of the discriminant 
\begin{equation}
\Delta =\hat{B}^{2}-4\hat{A}\hat{C}.  \label{del}
\end{equation}%

For the case $\Delta =0,$ there exist simple exact power-law solutions 
\begin{align}
a(\tau )& \propto \tau ^{\frac{2(2-\gamma )\omega +2+2c_{2}}{3\gamma
^{2}\omega -6\gamma (1+2\omega )+2(5+6\omega )-2c_{2}^{2}}}  \label{powera}
\\
\phi (\tau )& \propto \tau ^{\frac{2(4-3\gamma )-4c_{2}}{3\gamma ^{2}\omega
-6\gamma (1+2\omega )+2(5+6\omega )-2c_{2}^{2}}}.  \label{powerphi}
\end{align}%
Substituting these power-law solutions into (\ref{rho}) we can obtain the
corresponding power-law form for $\rho $ 
\begin{equation*}
\rho \sim \tau ^{\frac{4c_{2}(2-3\gamma -c_{2})-6\gamma (1+(2-\gamma )\omega
)}{3\gamma ^{2}\omega -6\gamma (1+2\omega )+2(5+6\omega )-2c_{2}^{2}}}.
\end{equation*}%
The relationship between $\tau $ and the cosmological time $t$ can now be
obtained by integrating the definition $d\tau =\rho a^{3}dt$ given in eq. (%
\ref{tau}). This gives (\ref{powera}) and (\ref{powerphi}) in terms of $t$ time as 
\begin{align}
a(t)& \sim t^{\frac{2+2(2-\gamma )\omega +2c_{2}}{4+3\gamma \omega (2-\gamma
)-2c_{2}(7-6\gamma -c_{2})}}  \label{power1} \\
\phi (t)& \sim t^{\frac{2(4-3\gamma )-4c_{2}}{4+3\gamma \omega (2-\gamma
)-2c_{2}(7-6\gamma -c_{2})}}. \label{power2}
\end{align}%
The condition required for the occurrence of power-law inflation is obtained
by requiring the power of time in equation (\ref{power2}) to exceed unity (for
the case with out energy transfer see refs. \cite{maeda, stein}).  For $\omega
>-3/2$ we always have $\Delta \geqslant 0,$ and the case $\Delta >0$
possesses the exact solutions 
\begin{align}
a(\tau )& =a_{0}(\hat{A}\tau ^{2}+\hat{B}\tau +\hat{C})^{\frac{(2-\gamma
)\omega +1+c_{2}}{2\hat{A}}}\left( \frac{2\hat{A}\tau +\hat{B}+\sqrt{\Delta }%
}{2\hat{A}\tau +\hat{B}-\sqrt{\Delta }}\right) ^{\frac{\hat{B}((2-\gamma
)\omega +1+c_{2})}{2\hat{A}\sqrt{\Delta }}}  \label{a+}
\\
\phi (\tau )& =\phi _{0}(\hat{A}\tau ^{2}+\hat{B}\tau +\hat{C})^{\frac{%
4-3\gamma -2c_{2}}{2\hat{A}}}\left( \frac{2\hat{A}\tau +\hat{B}+\sqrt{\Delta 
}}{2\hat{A}\tau +\hat{B}-\sqrt{\Delta }}\right) ^{\frac{(4-3\gamma -2c_{2})%
\hat{B}-\tau _{2}\hat{A}}{2\hat{A}\sqrt{\Delta }}}.  \label{phi+}
\end{align}%
where $a_{0}$ and $\phi _{0}$ are constants of integration.  For
$\omega <-3/2$ we have $\Delta \leqslant 0$, and the case $\Delta <0$ has the exact
solutions 
\begin{align}
a(\tau )& =a_{0}(\hat{A}\tau ^{2}+\hat{B}\tau +\hat{C})^{\frac{(2-\gamma
)\omega +1+c_{2}}{2\hat{A}}}\exp \left[ -\frac{((2-\gamma )\omega +1+c_{2})%
\hat{B}}{\hat{A}\sqrt{-\Delta }}\tan ^{-1}\left( \frac{\hat{B}+2\hat{A}\tau 
}{\sqrt{-\Delta }}\right) \right]   \label{a-}
\\
\phi (\tau )& =\phi _{0}(\hat{A}\tau ^{2}+\hat{B}\tau +\hat{C})^{\frac{%
4-3\gamma -2c_{2}}{2\hat{A}}}\exp \left[ \frac{2\tau _{2}\hat{A}-2(4-3\gamma
-2c_{2})\hat{B}}{\hat{A}\sqrt{-\Delta }}\tan ^{-1}\left( \frac{\hat{B}+2\hat{%
A}\tau }{\sqrt{-\Delta }}\right) \right] .  \label{phi-}
\end{align}%
These solutions have the same functional form as those found by Gurevich,
Finkelstein and Ruban \cite{Gur73} for Brans-Dicke theory, in the
absence of energy exchange ($\lambda =$ constant), and reduce to them in the limit $%
c_{2}\rightarrow 0$. The behaviour of these solutions at early and late
times will be discussed in the next section.

\subsection{$\protect\lambda (\protect\tau)=c_3 \protect\tau^n$, $n \neq 1$}

We now consider forms of $\lambda (\tau )$ that are more general than a
simple linear function of $\tau $. Making the choice $\lambda =c_{3}\tau
^{n} $, where $n\neq 1$ and $c_{3}$ is constant, and setting the free scalar
component to zero ($\tau _{2}=0$), we find that (\ref{phi'2}) and (\ref{V'2}%
) can be integrated exactly. The form of the solutions again depends upon
the roots of the denominator. For real roots we require $\omega >-3/2$, for
which we find the solutions 
\begin{align}
a(\tau )& =a_{0}\tau ^{\frac{2+2(2-\gamma )\omega }{\kappa }}\left[ \pm
2c_{3}\tau ^{n-1}\mp (4-3\gamma )\pm (2-\gamma )\sqrt{3(3+2\omega )}\right]
^{-\frac{3+3(2-\gamma )\omega -\sqrt{3(3+2\omega )}}{3\kappa (n-1)}}
\label{na+} \\
& \qquad \qquad \qquad \qquad \times \left[ \pm 2c_{3}\tau ^{n-1}\mp
(4-3\gamma )\mp (2-\gamma )\sqrt{3(3+2\omega )}\right] ^{-\frac{3+3(2-\gamma
)\omega +\sqrt{3(3+2\omega )}}{3\kappa (n-1)}}  \notag
\\
\phi (\tau )& =\phi _{0}\tau ^{\frac{2(4-3\gamma )}{\kappa }}\left[ \pm
2c_{3}\tau ^{n-1}\mp (4-3\gamma )\pm (2-\gamma )\sqrt{3(3+2\omega )}\right]
^{-\frac{(4-3\gamma )+(2-\gamma )\sqrt{3(3+2\omega )}}{\kappa (n-1)}}
\label{nphi+} \\
& \qquad \qquad \qquad \qquad \times \left[ \pm 2c_{3}\tau ^{n-1}\mp
(4-3\gamma )\mp (2-\gamma )\sqrt{3(3+2\omega )}\right] ^{-\frac{(4-3\gamma
)-(2-\gamma )\sqrt{3(3+2\omega )}}{\kappa (n-1)}}  \notag
\end{align}%
where $\kappa \equiv 2(5-3\gamma )+3(2-\gamma )^{2}\omega $. For a
denominator with imaginary roots we require $\omega <-3/2$, for which we
find 
\begin{align}
a(\tau )& =a_{0}\left[ \pm 2(4-3\gamma )c_{3}\tau ^{1-n}\pm \kappa \tau
^{2(1-n)}\mp 2c_{3}^{2}\right] ^{\frac{1+(2-\gamma )\omega }{\kappa (1-n)}}
\label{na-} \\
& \qquad \qquad \qquad \qquad \times \exp \left\{ \frac{2\sqrt{-3(3+2\omega )%
}}{3\kappa (1-n)}\tan ^{-1}\left( \frac{(4-3\gamma )-2c_{3}t^{n-1}}{%
(2-\gamma )\sqrt{-3(3+2\omega )}}\right) \right\}   \notag 
\\
\phi (\tau )& =\phi _{0}\left[ \pm 2(4-3\gamma )c_{3}\tau ^{1-n}\pm \kappa
\tau ^{2(1-n)}\mp 2c_{3}^{2}\right] ^{\frac{(4-3\gamma )}{\kappa (1-n)}}
\label{nphi-} \\
& \qquad \qquad \qquad \qquad \times \exp \left\{ -\frac{2(2-\gamma )\sqrt{%
-3(3+2\omega )}}{\kappa (1-n)}\tan ^{-1}\left( \frac{(4-3\gamma
)-2c_{3}t^{n-1}}{(2-\gamma )\sqrt{-3(3+2\omega )}}\right) \right\}   \notag
\end{align}%
with $\kappa $ defined as above. The $\pm $ and $\mp $ signs here indicate
that there are multiple solutions that satisfy the field equations. These
signs should be chosen consistently within each set of square brackets
(solutions (\ref{na+}) and (\ref{nphi+}) can have upper or lower branches
chosen independently in each set of square brackets, as long as a consistent
branch is taken within each separate set of square brackets). The physical
branch should be chosen as the one for which the quantity in brackets
remains positive as $\tau \rightarrow \infty $, so ensuring the existence of
a positive real root in this limit,

These solutions display interesting new behaviours at both early and late
times, which will be discussed in the next section.

\section{Behaviour of Solutions}

These exact solutions for $a(\tau )$ and $\phi (\tau )$ can now be analysed
at early and late times.

\subsection{$\protect\lambda (\protect\tau)= c_1+c_2 \protect\tau$}

At late times, as $\tau \rightarrow \infty $, the solutions (\ref{a+})-(\ref%
{phi-}) all approach the\ exact power-law solutions (\ref{powera}) and (\ref%
{powerphi}). It can be seen that these solutions reduce to the usual
spatially flat FRW Brans-Dicke power-law solutions \cite{Nar} in the limit
that the rate of energy transfer goes to zero, $c_{2}\rightarrow 0$. It can
also be seen that these solutions reduce to the spatially-flat FRW general
relativistic solutions in the limit $\omega \rightarrow \infty $,
irrespective of any (finite) amount of energy transfer.

The early-time behaviour of these solutions approaches that of the general
Brans-Dicke solutions \cite{Gur73}, without energy transfer. Generally, we
expect an early period of free-scalar-field domination except in the case $%
\tau _{2}=0$, in which case the power-law solutions (\ref{powera}) and
(\ref{powerphi}) are valid right up
to the initial singularity. For $\omega >-3/2,$ the scalar-field dominated
phase causes an early period of power-law inflation. In this case there is
always an initial singularity and the value of the scalar field diverges to
infinity or zero as it is approached, depending on the sign of $\tau _{2}$.
For $\omega <-3/2$ there is a `bounce' and the scale factor has a minimum
non-zero value. In these universes there is a phase of contraction followed
by a phase of expansion, with no singularity separating them. Solutions of
this type were the focus of \cite{Bar04} where the evolution of $\phi $
through the bounce was used to model the variation of various physical
constants in such situations. The energy exchange term does not play a
significant role at early times in these models. The asymptotic solutions as
the singularity (or bounce) is approached are the same as if the energy
exchange term had been neglected, and are given by \cite{Tup}, up to the
absorption of $c_{1}$ into $\tau _{2}$ previously described.

\subsection{$\protect\lambda (\protect\tau)=c_3 \protect\tau^n$, $n \neq 1$}

The behaviour of the solutions (\ref{na+})-(\ref{nphi-}) depends upon the
signs of $n-1$ and $c_{3},$ as well as on the sign of $\omega +3/2$. For
illustrative purposes we will consider the radiation case $\gamma =4/3$
which is appropriate for realistic universes dominated by
asymptotically-free interactions at early times.

For $n>1$ it can be seen that the late-time attractors of solutions (\ref%
{na+})-(\ref{nphi-}) are $a\rightarrow $ constant and $\phi \rightarrow $
constant as $\tau \rightarrow \infty $, for both $\omega >-3/2$ and $\omega
<-3/2$. At late times these universes are asymptotically static; the
evolution of the scale-factor ceases as $\tau \rightarrow \infty $ and both $%
\phi $ and $\rho $ become constant. Further analysis is required to
establish whether these static universes are stable or not (we expect them
to be stable as no tuning of parameters or initial conditions has been
performed to obtain these solutions).

\begin{figure}[p]
\mbox{\subfigure[$c_3=10$]{\epsfig{figure=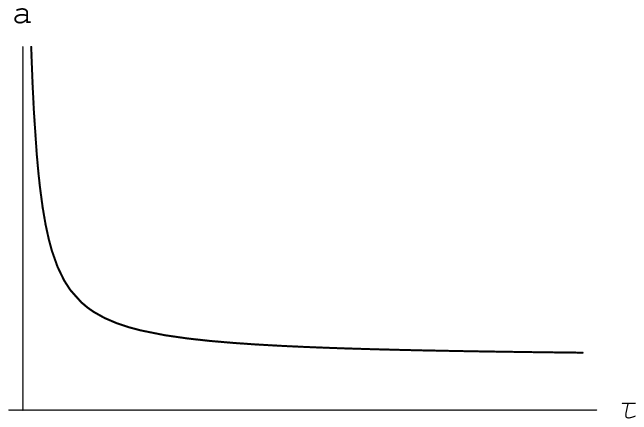,height=3.5cm}\label{a}}
\quad
\subfigure[$c_3=-10$]{\epsfig{file=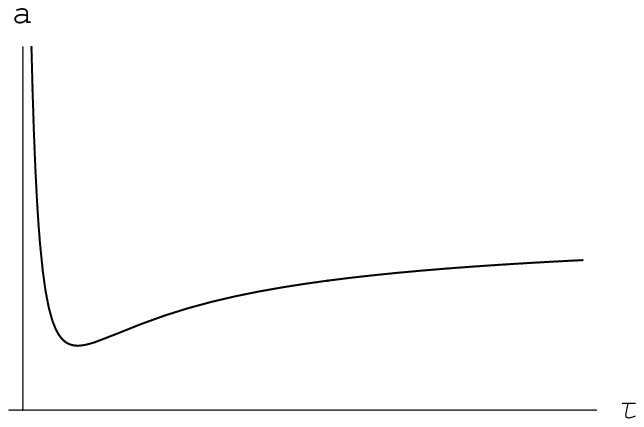,height=3.5cm}\label{b}}} 
\mbox{\subfigure[$c_3=10$]{\epsfig{figure=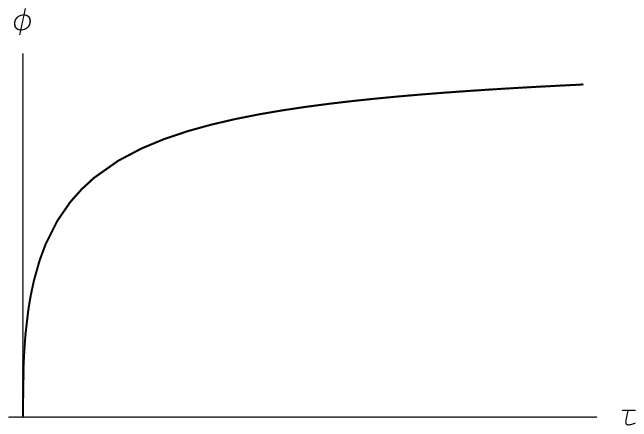,height=3.5cm}\label{c}}
\quad
\subfigure[$c_3=-10$]{\epsfig{file=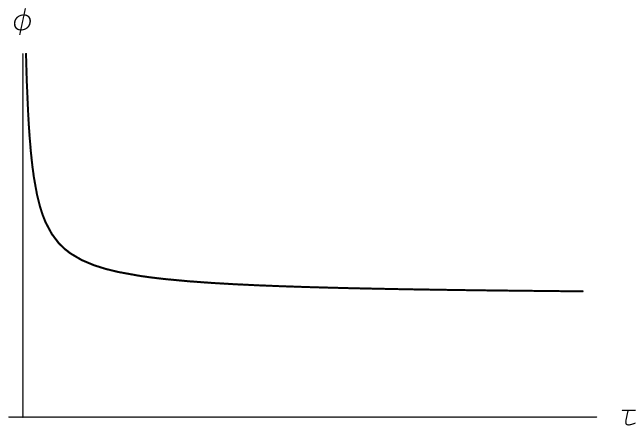,height=3.5cm}\label{d}}}
\caption{The time evolution of $a$ and $\protect\phi $ for $n=2$, $\protect%
\omega =10$ and $\protect\gamma =4/3$.}
\end{figure}

The early-time behaviour of solutions with $n>1$ depends upon the sign of $%
c_{3}$ as well as whether $\omega $ is greater or less than $-3/2$. We will
consider first the case of $\omega >-3/2$. For $c_{3}>0$, we see that $%
a\rightarrow \infty $ as $\tau ^{n-1}\rightarrow \tau _{0}^{+};$ for $%
c_{3}<0 $ we see that $a\rightarrow \infty $ as $\tau ^{n-1}\rightarrow \tau
_{0}^{-} $ (where $\tau _{0}^{+}=((2-\gamma )\sqrt{3(3+2\omega )}+(4-3\gamma
))/2c_{3} $ and $\tau _{0}^{-}=-((2-\gamma )\sqrt{3(3+2\omega )}-(4-3\gamma
))/2c_{3}$). For $n>1$ and $\omega >-3/2$ we therefore have the generic
behaviour that $a\rightarrow \infty $ at early times and $a\rightarrow $
constant at late times. The behaviour of $a$ at intermediate times varies in
form depending on the sign of $c_{3}$, as can be seen in Figures \ref{a} and %
\ref{b}. The asymptotic form of $\phi $ for $n>1$ and $\omega >-3/2$ depends
critically on the sign of $c_{3}$. For $c_{3}>0$ it can be seen that $\phi
\rightarrow 0 $ as $\tau ^{n-1}\rightarrow \tau _{0}^{+}$, whereas for $%
c_{3}<0$ it can be seen that $\phi \rightarrow \infty $ as $\tau
^{n-1}\rightarrow \tau _{0}^{-} $. The behaviour of $\phi $ in these two
cases is illustrated in Figures \ref{c} and \ref{d}.

\begin{figure}[p]
\mbox{\subfigure[$c_3=10$]{\epsfig{figure=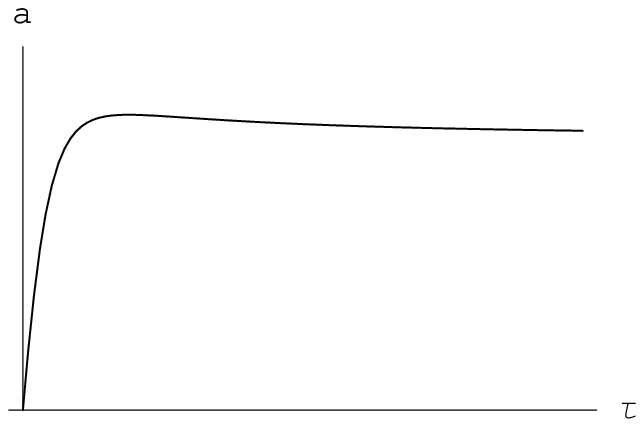,height=3.5cm}\label{e}}
\quad
\subfigure[$c_3=-10$]{\epsfig{file=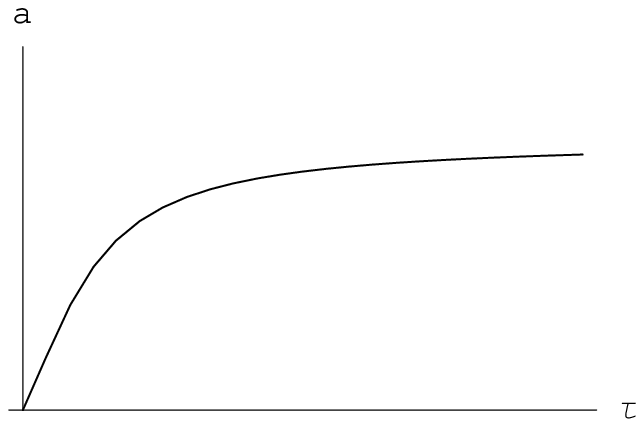,height=3.5cm}\label{f}}} 
\mbox{\subfigure[$c_3=10$]{\epsfig{figure=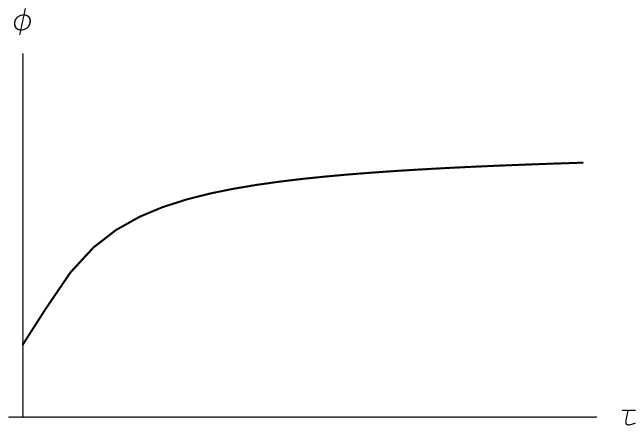,height=3.5cm}\label{g}}
\quad
\subfigure[$c_3=-10$]{\epsfig{file=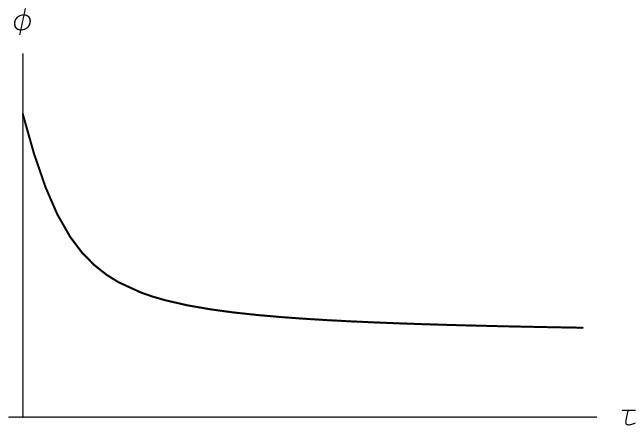,height=3.5cm}\label{h}}}
\caption{The time evolution of $a$ and $\protect\phi $ for $n=2$, $\protect%
\omega =-10$ and $\protect\gamma =4/3$.}
\end{figure}

We now consider the early-time behaviour of solutions with $n>1$ and $\omega
<-3/2$. It can be seen that $a\rightarrow 0$ as $\tau \rightarrow 0$
irrespective of the sign of $c_{3}$, so that we find the generic behaviour $%
a\rightarrow 0$ at early times and $a\rightarrow $ constant at late times
(this is in contrast to the standard theory where an initial singularity is
avoided when $\omega <-3/2$). Again, the behaviour of $a$ at intermediate
times is dependent on the sign of $c_{3}$, as can be seen from Figures \ref%
{e} and \ref{f}. As $\tau \rightarrow 0$ we see that $\phi $ has a finite
non-zero value and is either increasing or decreasing depending on the sign
of $c_{3}$. This behaviour is shown in Figures \ref{g} and \ref{h}. 

\begin{figure}[p]
\mbox{\subfigure[$c_3=10$]{\epsfig{figure=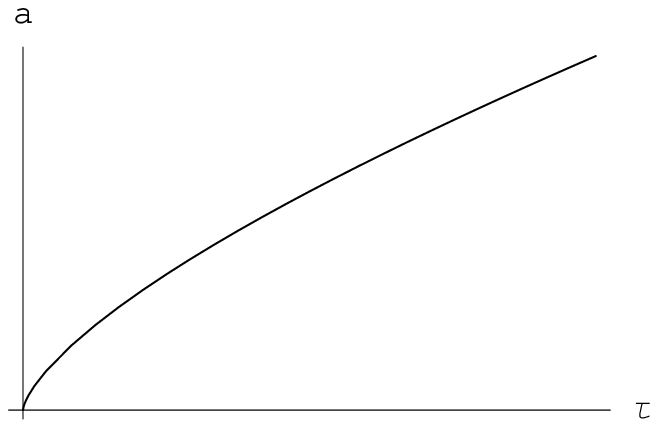,height=3.5cm}\label{i}}
\quad
\subfigure[$c_3=-10$]{\epsfig{file=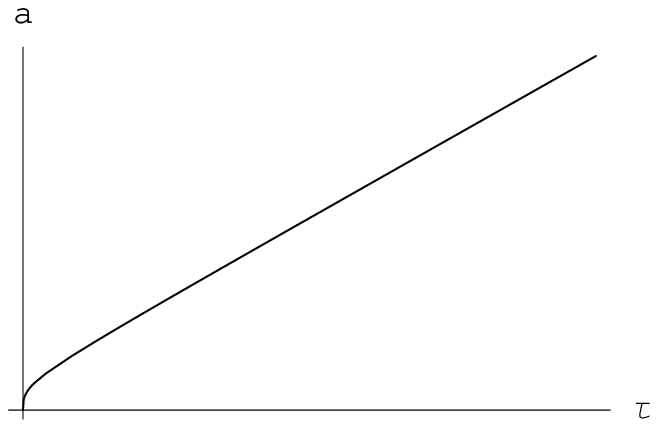,height=3.5cm}\label{j}}} 
\mbox{\subfigure[$c_3=10$]{\epsfig{figure=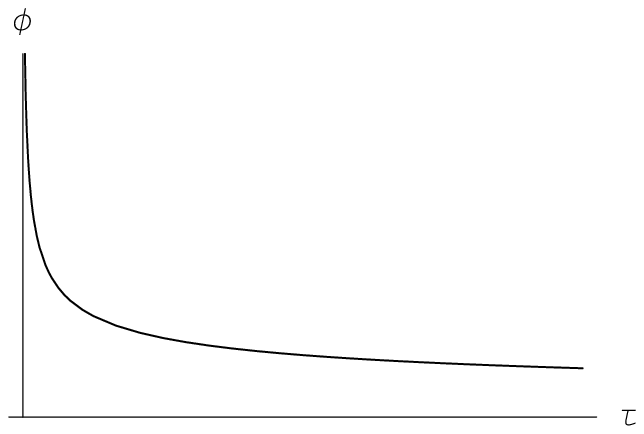,height=3.5cm}\label{k}}
\quad
\subfigure[$c_3=-10$]{\epsfig{file=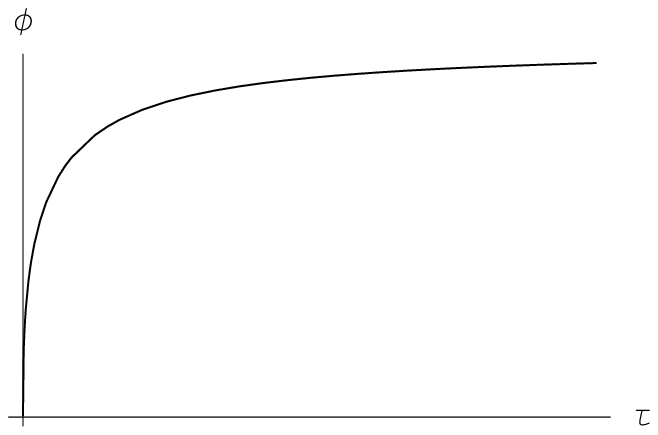,height=3.5cm}\label{l}}}
\caption{The time evolution of $a$ and $\protect\phi $ for $n=0$, $\protect%
\omega =10$ and $\protect\gamma =4/3$.}
\end{figure}
\begin{figure}[p]
\mbox{\subfigure[$c_3=10$]{\epsfig{figure=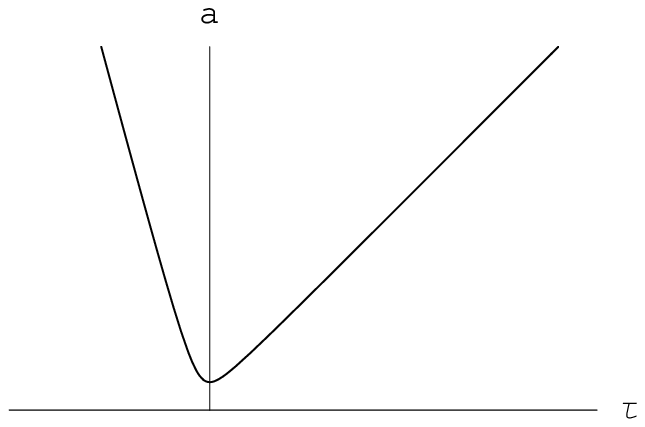,height=3.5cm}\label{m}}
\quad
\subfigure[$c_3=-10$]{\epsfig{file=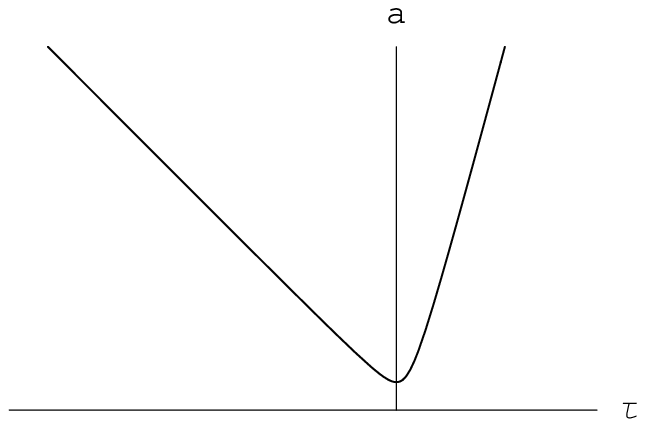,height=3.5cm}\label{n}}} 
\mbox{\subfigure[$c_3=10$]{\epsfig{figure=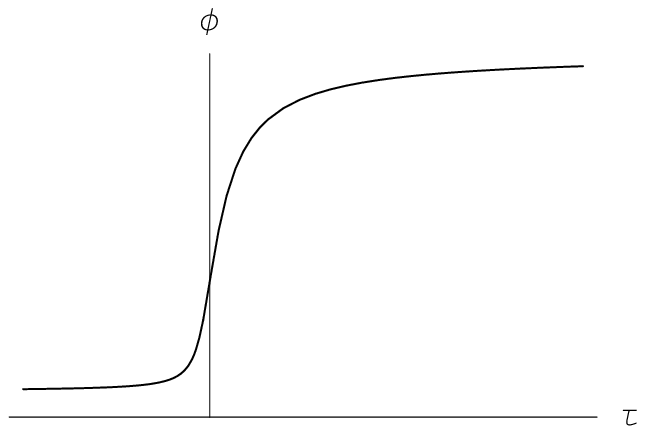,height=3.5cm}\label{o}}
\quad
\subfigure[$c_3=-10$]{\epsfig{file=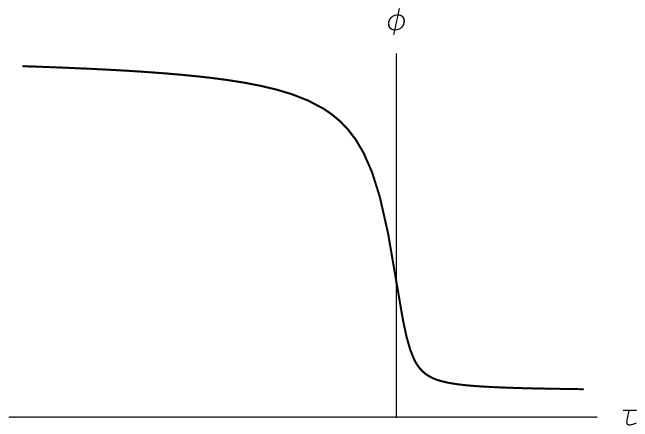,height=3.5cm}\label{p}}}
\caption{The time evolution of $a$ and $\protect\phi $ for $n=0$, $\protect%
\omega =-10$ and $\protect\gamma =4/3$.}
\end{figure}

It remains to investigate the nature of the solutions with $n<1$. At late
times we see that 
\begin{align*}
a& \rightarrow \tau ^{\frac{2+2(2-\gamma )\omega }{2(5-3\gamma )+3(2-\gamma
)^{2}\omega }}, \\
\phi & \rightarrow \tau ^{\frac{2(4-3\gamma )}{2(5-3\gamma )+3(2-\gamma
)^{2}\omega }},
\end{align*}%
as $\tau \rightarrow \infty $, irrespective of the sign of $c_{3}$ or the
value of $\omega $. These late-time attractors are the flat FRW power-law
Brans-Dicke solutions \cite{Nar} which reduce to the standard
general-relativistic solutions in the limit $\omega \rightarrow \infty $.

The early-time behaviour when $n<1$ depends on the sign of $c_{3}$ and the
sign of $\omega +3/2$. We consider first the case $\omega >-3/2$. In this
case it can be seen that $a\rightarrow 0$ as $\tau ^{n-1}\rightarrow \tau
_{0}^{+}$ or $\tau ^{n-1}\rightarrow \tau _{0}^{-}$, for $c_{3}>0$ or $%
c_{3}<0$ respectively. The behaviour of $a$ for both $c_{3}>0$ and $c_{3}<0$
is shown in Figure \ref{i} and \ref{j}. The behaviour of $\phi $ at early
times depends on the sign of $c_{3}$ and goes to $\infty $ for $c_{3}>0$ or
to $0$ for $c_{3}<0$. The behaviour of $\phi $ in these cases is shown in
Figures \ref{k} and \ref{l}. For $\omega <-3/2$ the scale factor $a$
contracts to a finite, but non-zero, minimum value and then expands. The
exact form of the minimum depends on the values of $n$, $\omega $ and $%
c_{3}, $ but it is interesting to note that odd values of $n$ produce
symmetric bounces and even values of $n$ produce asymmetric bounces, as
illustrated in Figures \ref{m} and \ref{n}. The evolution of $\phi $ through
these bounces is smooth with a time direction prescribed by the value of $%
c_{3}$, as shown in Figures \ref{o} and \ref{p} (increasing for $c_{3}>0$
and decreasing for $c_{3}<0$). The effect of changing the sign of $c_{3}$ is
seen to be a mirroring of the evolution of $a$ and $\phi $ in the $y-$axis. 

\section{Physical Consequences}

The solutions found in the previous sections are of physical interest
for a number of reasons.  The transfer of energy and momentum between a
non-minimally coupled scalar field $\phi$ and matter fields is a
prediction of a number of fundamental theories of current interest,
including string theories, Kaluza-Klein theories and brane-worlds.  The
cosmologies produced by such an interaction, therefore, should be of
direct interest in the consideration of these theories.  Furthermore,
the solutions we have found display modified behaviour at both early
and late times.  The investigation of modified theories of gravity at
early times is of particular interest as it is in the high-energy limit that
deviations from general relativity are usually expected.  Modified
behaviour at late times is also of interest as it is
at these times that we can make direct observations which can be used
to constrain deviations from the standard
general-relativistic model.  We will now summarise the behaviour
of the solutions found in the previous section, highlighting
the physically significant results and constraints that can be placed
on the theory by observations.

For the case of $\lambda $ linear in $\tau $ it was shown that the late-time
attractors of the general solutions are no longer the power-law solutions of
the Brans-Dicke theory, but are given by equations (\ref{power1}) and (\ref%
{power2}).  These attractors are of special interest as they have a
simple power-law form that reduces to the general relativistic result
in the limit $\omega \rightarrow \infty$ and to the Brans-Dicke result
in the limit $c_2 \rightarrow 0$.  Observations of cosmic microwave background anisotropies and
the products of primordial nucleosynthesis will therefore be able to constraint any
potential late-time deviations of this kind, and hence the underlying
model.  The process of primordial nucleosynthesis in scalar-tensor theories has been
used by a number of authors to place constraints on the coupling
parameter $\omega (\phi )$ \cite{Cas1, Cas2, Ser1, Ser2, San, Dam, Cli}.  In these
studies the different value of $G$ during nucleosynthesis causes the
weak interactions to freeze out at a different time and hence the
proton to neutron ratio at this time is different to the standard
case.  This modification causes different abundances of the light
elements to be produced, which can be compared with observations to
constrain the underlying theory.  Studies of this kind usually assume
$G$ to be constant during nucleosynthesis, which will not be the case
when energy is allowed to be exchanged between $\phi$ and the matter
fields.  The effects of a non-constant $G$ were studied in
\cite{Cli}.  A similar study would be required to place constraints
upon the parameters $c_2$ and $\omega$ in this theory.  The
cosmic microwave background power spectrum has also often been used to constrain
scalar-tensor theories of gravity \cite{cmb1, cmb2, cmb3, cmb4, cmb5,
  cmb6}.  In these studies the redshift of matter-radiation equality
is different from its usual general relativistic value due to
the modified late-time evolution of the Universe.  This change in the
redshift of equality is imprinted on the spectrum of perturbations
as it is only after equality that
sub-horizon scale perturbations are allowed to grow.  The main effect
is seen as a shift in the first peak of the power-spectrum, which can be
compared with observations to constrain the theory.  Again, the late
time evolution of the Universe is modified from the usual Brans-Dicke
case by the energy exchange that we consider, so that the previous
constraints are not directly applicable.

For the case of a non-linear power-law exchange of energy, described by $%
\lambda \propto \tau ^{n}$, the late-time evolution of $a$ and $\phi $ can
be significantly modified. For $n>1,$ the solutions do not continue to
expand eternally, but are attracted towards a static state where the
time-evolutions of $a$, $\phi $ and $\rho $ cease. For $n<1$ the generic
late-time attractor is the power-law solution of a flat FRW Brans-Dicke
universe. It appears that theories of energy exchange with $n>1$ are ruled
out immediately by observations of an expanding universe whilst the case of $%
n<1$ is subject to the same late-time constraints as the standard
Brans-Dicke theory \cite{Cas1, Cas2, Ser1, Ser2, San, Dam, Cli,cmb1,
  cmb2, cmb3, cmb4, cmb5, cmb6}.

It remains to investigate the physical consequences of the early-time
behaviour of our solutions.  For $\lambda $ linear in $\tau $ these solutions
approach those of the standard Brans-Dicke theory as either the initial
singularity or the minimum of the bounce are approached, according to the
sign of $\omega -3/2$.  The physical significance of this behaviour
has been discussed many times before, usually focusing on the
avoidance of the initial singularity and the inflation that can result
from the presence of the free component of the scalar field.

\begin{figure}[tbp]
\mbox{\subfigure{\epsfig{figure=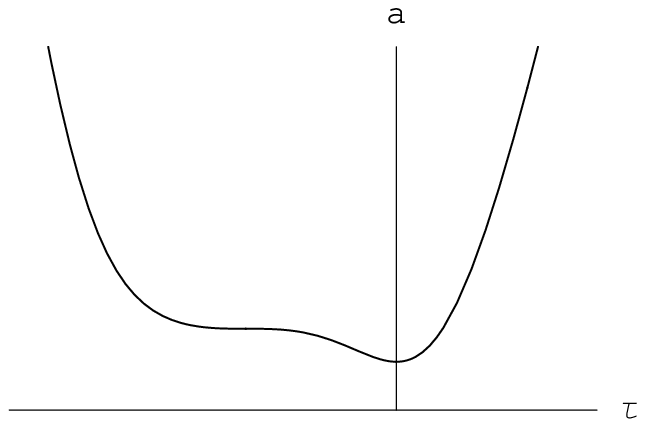,height=3.5cm}}
\quad
\subfigure{\epsfig{file=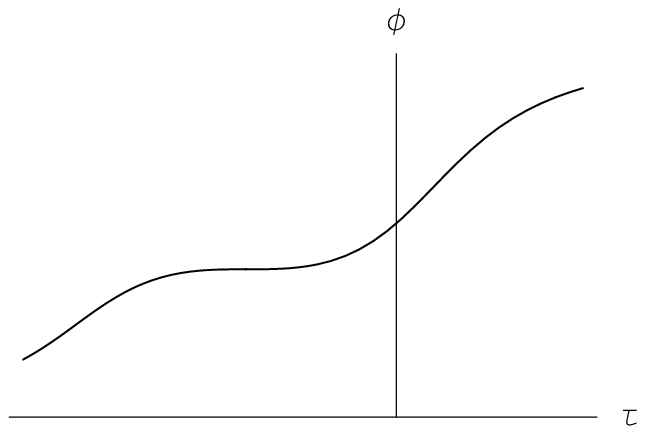,height=3.5cm}}}
\caption{The time evolution of $a$ and $\protect\phi $ for $n=-2$, $\protect%
\omega =-10$, $c_{3}=10$ and $\protect\gamma =4/3$.}
\label{bounce}
\end{figure}

For $\lambda \propto \tau ^{n}$ the early-time
behaviour can be significantly changed from that of the standard theory. For 
$n>1$ the scale factor $a$ either approaches infinity or zero, depending on
our choice of $\omega $ and $c_{3}$, as previously described. For the more
realistic case of $n<1$ the evolution of $a$ at early times either undergoes
a period of rapid expansion or a non-singular bounce, depending on whether $%
\omega $ is greater or less than $-3/2$. This behaviour is similar to that
of the general solutions of the standard theory, but in this case the free
scalar-field-dominated epoch has not been invoked and there is more freedom
as to the exact form of the evolution. For example, with a suitable choice
of parameters it is possible to create a universe that contracts and then is
briefly static before `bouncing' and continuing on to its late-time
power-law evolution. This is shown in Figure \ref{bounce} for the case $%
\omega =-10$, $n=-2$ and $c_{3}=10$. (It is interesting to note that Peter
and Pinto-Neto remark that a static period followed by a bounce could
potentially produce a scale-invariant spectrum of perturbations \cite{Peter}%
). 

For the physically reasonable models with $n\leqslant 1$ the evolution of $%
\phi $, and hence of $G$, can be significantly altered at early times from
what is generally assumed to be the case in scalar-tensor theories of
gravity. For the case $\omega >-3/2,$ the value of $G$ can be made to
diverge to infinity or to zero as the initial singularity is approached,
independent of whether or not there was an early scalar-dominated phase to
the universe's history. For the case $\omega <-3/2,$ the value of $G$
evolves smoothly through the bounce in the scale-factor, and is again
independent of whether or not there was a scalar dominated phase. More
complicated evolutions of $\phi $ can also be constructed, as can be seen
in Figure \ref{bounce}.

\section{Discussion}

We have considered spatially-flat FRW universes in scalar-tensor theories of
gravity where energy is allowed to be exchanged between the Brans-Dicke
scalar field that determines the strength of gravity and any perfect-fluid
matter fields in the space-time. We have presented a prescription for
integrating the field equations exactly for some unknown function $\lambda $
which describes the rate at which energy is exchanged. For the case of $%
\lambda $ being a linear function of $\tau $ we have found the general
solutions to the problem and for the case of $\lambda $ being a non-linear
power law function of $\tau $ we have been able to find a wide class of
exact solutions. These solutions display behaviours that can deviate
substantially from the corresponding solutions in the standard case, where
the exchange of energy is absent. Depending upon the values of the
parameters defining the theory and the exchange of energy, deviations in the
evolution of $a$ and $\phi $ can occur at both early and late times,
providing a richer phenomenology than is available in the standard
theory.  

We have found that the parameter $n$ must be bounded by the
inequality $n \leqslant 1$ if the Universe is to be expanding at late
times.  For $n=1$ we have found late-time power-law attractor
solutions which can be used to constrain the parameters $c_2$ and
$\omega$.  For $n <1$ we have seen that the late-time evolution will
be that same as in the standard Brans-Dicke case, and so is subject
to the same observational constraints as these theories.  The
parameters $c_3$ and $\tau_2$ have been shown to be influential only
in the vicinity of the initial singularity, or at the minimum of expansion in
non-singular solutions.  These parameters are therefore less
accessible to constraint by late-time observations (see, however, \cite{Cli} where
the influence on primordial nucleosynthesis is used to constrain $\tau_2$).  The parameter
$\omega$ is, as always, subject to the very tight solar system
constraint $\omega > 40 000$ to $2 \sigma$ \cite{Bert}.

These results could be of interest in
attempting to explain why $G$ is so small in the present day Universe
compared to the proton mass scale ($Gm_{pr}^{2}\sim 10^{-39}$).  In
these models the value of $G$ can decay away by a coupling between the
scalar field $\phi$ and the matter fields which allows energy to be
transferred.  The small value of $G$ is then due to the age of the
Universe.  It remains to see
whether or not the late-time modifications found above are consistent
with observations of the primordial abundance of light elements,
microwave background formation and other late-time physical
processes.  Whilst being beyond the scope of this article, these
studies should be able to be performed in an analogous way to the ones
that already exist for the standard Brans-Dicke theory.

Using the late time solutions that have been found it is possible to comment
on the case of FRW cosmologies with non-zero spatial curvature. At early
times it is expected that the effect of any spatial curvature on the
evolution of $a(t)$ should be negligible. From the solution (\ref{powera})
we can see that spatial curvature will dominate the late time evolution if
the condition 
\begin{equation*}
\frac{2+2(2-\gamma )\omega +2c_{2}}{4+3\gamma \omega (2-\gamma
)-2c_{2}(7-6\gamma -c_{2})}<1
\end{equation*}%
is satisfied. If this condition is not satisfied then the power-law solution
(\ref{powera}) will be an attractor as $t\rightarrow \infty $ even in the
case of non-zero spatial curvature, offering a potential solution to the
flatness problem. This behaviour corresponds to power-law inflation
and it can be seen that the condition for equation (\ref{powera}) to dominate over the spatial curvature
at late times is, indeed, also the condition that power-law inflation should
occur.

In conclusion, we have found that a direct coupling between $\phi $ and the
matter fields in scalar-tensor cosmologies provides a richer frame-work
within which one can consider variations of $G$. We have shown that it is
possible to construct models where the late-time violations of the
equivalence principle can be made arbitrarily small (for $\lambda \propto
\tau $) or are attracted to zero (for $\lambda \propto \tau ^{n}$ where $n<1$%
). This enlarged phenomenology is of interest for the consideration of the
four-dimensional cosmologies associated with higher-dimensional theories as
well as for more general considerations of the variation of $G$ and its
late-time value. This study has been limited to scalar-tensor theories with
constant coupling parameters, to flat FRW cosmologies, and to special cases
of $\sigma ^{0}$ that allow direct integration of the field equations.
Obvious extensions exist in which these assumptions are partially or
completely relaxed.

\noindent\newline

{\large \textbf{ACKNOWLEDGEMENTS}}\newline

T. Clifton is supported by the PPARC.

\end{document}